\documentstyle[aps,prl,multicol]{revtex}
\def\beq{\begin{equation}}
\def\eeq{\end{equation}}
\begin{document}                
\title{Anomalous flux-flow dynamics in layered type-II superconductors at
low temperatures}
\author{M. V. Feigel'man and M. A. Skvortsov}
\address{L. D. Landau Institute for Theoretical Physics, Moscow 117940, RUSSIA}
\maketitle

\begin{abstract}
Low-temperature dissipation due to vortex motion
in strongly anisotropic type-II
superconductors with a moderate disorder
($\Delta^2/E_F \ll \hbar/\tau \ll \Delta$)
is shown to be determined
by the Zener-type transitions between the
localized electronic states in the vortex core.
Statistics of these levels
is described by the random matrix ensemble of the class C
defined recently by Atland and Zirnbauer \cite{AZ},
so the vortex motion leads naturally
to the new example of a parametric statistics of energy levels.
The flux-flow conductivity $\sigma_{xx}$
is a bit lower than the quasiclassical one \cite{KK} and
{\it grows} slowly with the increase of the electric field.
\end{abstract}

\begin{multicols}{2}

It is generally accepted that the flux-flow longitudinal
($\sigma_{xx}$) and Hall ($\sigma_{xy}$) conductivities in the mixed
state of type-II superconductors are given, respectively,
by the Bardeen-Steven (BS) \cite{BS}
and Nozieres-Vinen \cite{NV} relations
\beq
\sigma_{xx} \approx \sigma_n \frac{H_{c2}}{B};
\quad \sigma_{xy} \approx {(\omega_0\tau)} \sigma_{xx},
\label{BS}
\eeq
where $\sigma_n $ is the normal-state conductivity,
$\tau$ is the elastic lifetime with respect to
impurity scattering,
and $\hbar\omega_0 \approx \Delta^2/E_F \ll \Delta$
(it is supposed in (\ref{BS}) that $\omega_0\tau \ll 1$).
The meaning of this relation is very simple: it reflects the
fact that the core region of a vortex
(of the area $\sim \xi^2$) may be considered
(with respect to its electronic properties) just as a normal metal.
This idea is based on the existence
of localized electronic levels within the core \cite{CdGM},
with energies constituting an equidistant
set $E_{\mu} = \mu\hbar\omega_0$,
where $\mu = i+1/2$ and $i$ is integer.
A microscopic theory of flux-flow conductivities,
which explicitly takes into account an existence of this
level structure was developed long ago \cite{KK}
(cf.~also \cite{kop95}, \cite{Pisma}, \cite{OtFeiGesh}).
This theory is based on the quasiclassical
nonequilibrium diagram technique (see e.g.\cite{LO});
its predictions for $\rho_{xx}$ and $\rho_{xy}$
basically coincide with the above simple picture
as far as the superconductor
is not in the so-called superclean limit (i.e. if the inverse
electron-impurities scattering time $\tau^{-1}$
is larger than the level separation $\omega_0$).
The underlying idea of this theory is
 that interaction between electrons
localized near the (moving) vortex
and impurities can be treated as a kind of scattering problem
in the continuous spectrum,
i.e. similar to the impurity scattering in a normal state.

In the present Letter we will show that in the case
of strongly anisotropic (layered) superconductors
the validity of the above picture is
limited to the situations when
inelastic width $\Gamma$ of the core electronic levels
is comparable to or large than the level spacing $\omega_0$.
Note that
$\Gamma = \Gamma_{int}+ \Gamma_v$, where $\Gamma_{int}$ is due
to electron-electron and (mainly) electron-phonon interactions
and grows with temperature,
whereas $\Gamma_v$ is due to nonstationarity of
the impurity potential around moving vortex,
and grows with its velocity $v_v$;
therefore the condition $\Gamma \gg \omega_0$ is fulfilled
at sufficiently large $T$ or $v_v$.
In the opposite limit, $\Gamma \ll \omega_0$, the core levels are
 essentially discrete;
as a result, all dissipation
induced by the motion of a vortex is due to rare nonadiabatic (Zener)
processes of electron excitations between the core levels
and subsequent downward electron transitions
with emission of phonons.
As a result, the longitudinal flux-flow conductivity
$\sigma_{xx}$ becomes reduced compared
to its ``quasiclassical" value (\ref{BS}),
and dependent on the electric field:
\beq
\sigma_{xx} \leq \frac{ecn_e}{B}\omega_0\tau
\left(1 - \sqrt{\frac{E_*}{E}}\right) ;
\quad E_* \approx \frac{\omega_0^2 B}{c}\sqrt{\frac{\hbar}{\rho s^3}},
\label{new}
\eeq
where the characteristic electric field $E_*$ is very small,
so in the domain of applicability of our theory
$E \gg E_*$ (here $\rho$ is the mass density
of the crystal, $s$ is some average sound velocity
and $c$ is the velocity of light).
Note that according to Eq.~(\ref{new}) conductivity
{\it grows} with the increase of electric field (or current),
which is the trend opposite to the one known near
the classical vortex depinning transition \cite{review}.
Our result (\ref{new}) is expected to be valid
in the range of not very high current densities
$j \leq j_K = j_0 (k_F l)^{-1/2}$,
where $l = v_F\tau$ is the elastic mean-free-path, and
$j_0 \sim e n_e\Delta/p_F$ is the depairing current density.
Note that the above result refers to the moderately clean case
$\omega_0 \ll \tau^{-1} \ll \Delta/\hbar$,
where the {\it average} density of states within the core
$\langle\nu_c(E)\rangle$ is constant (at $E \geq \hbar\omega_0$)
as in normal metal.
Therefore the effects we are discussing are of
{\it mesoscopic nature} in the sense that they
``feel" the structure of correlations of energy levels.
Thus the range of existence of these
effects disappears with the growth of $k_Fl$:
$j_K \propto (k_Fl)^{-1/2}$.

We are going to consider strongly anisotropic superconductors with
a high effective mass anisotropy $m_c/m_a \gg1$; it means
very weak dispersion of the core level's energies
as a function of $k_z$,
$(E_{\mu}(k_z)-E_{\mu}(0))/\hbar \sim %
\mu\omega_0 \frac{\mathstrut m_a}{m_c} %
\frac{\vphantom{A^A_A} k_z^2}{\vphantom{A^A_A} k_F^2}\ll \omega_0$.
It allows, as long as the low-lying levels
with $\mu \ll m_c/m_a$ are relevant,
to treat the core levels as purely discrete,
which formally corresponds to a ``pancake" vortex
in 2D superconductor
(one important point of our arguments where 3D nature of vortex
is essential will be explained later).

At low temperatures any dissipation of energy
related to the vortex motion is due to transitions
between discrete electron states within the core.
In the absence of electron-phonon interaction these levels have almost
zero width (inelastic width due to electron-electron
interactions for the system of discrete levels
at low temperatures is even much
weaker than the electron-phonon one, cf.~\cite{blanter}),
whereas their exact locations depend
on the particular realization of impurities
in the region of the size $\sim \xi$
around the vortex center ${\bf r}_v$.
The most direct  way to study the statistics of random energy
levels is to employ powerful machinery
of the Supersymmetric Sigma-Model \cite{efetov}.
Here, however, we prefer to use technically simpler
(although less general) path,
sufficient for the calculation of dissipation rate in the
case $\omega_0\tau \ll 1$.
This condition ensures that the position of each level $E_i$
is strongly modified with respect to the bare levels \cite{CdGM}
of an ideally clean superconductor.
The second important point is that the wavefunctions
corresponding to all these levels
are confined in the same area $\sim\xi^2$ within
the vortex core.
Therefore it is quite natural to describe the distribution
of levels in terms of an appropriate
Random Matrix Ensemble (RME) \cite{RMT}.
When vortex is moving, the realization
of disorder in the core region is changing,
so the core levels are moving up and down,
presenting a new example of a ``parametric level statistics"
studied before with regard to disordered metallic grains
and quantum dots (see e.g. \cite{AltSai}).
In our case the parameter $X(t)$ governing the evolution
of levels is just the vortex coordinate,
so one can try to make use of the
relation \cite{Wilk1,Wilk2,AltSai}
between the rate of energy dissipation per pancake vortex
(within the applicability of the standard Kubo approach)
and the mean-squared level
``velocities" $\langle(dE_i/d X)^2\rangle$:
\beq
\left(\partial W/\partial t\right)_{Kubo}
    = \frac{\beta}{2}\pi\hbar C(0) (\partial X/\partial t)^2,
\label{C0}
\eeq
where $\beta = 1,2,4$ for the system described by,
respectively, orthogonal, unitary and symplectic
Wigner-Dyson ensembles, and $C(0)$ is the normalized
dispersion of level ``velocities",
$C(0) = \langle(dE_i/d X)^2\rangle/(\hbar\omega_0)^2$
(here $\hbar\omega_0 = \langle E_{i+1}-E_i \rangle$).
Vortex solution breaks T-invariance,
whereas leaves intact invariance with
respect to spin rotations
(here and below we neglect Zeeman spin splitting, which is
weak as long as  $B \ll H_{c2}$).
Therefore one could, at first sight,
conclude that the relevant for our problem
Wigner-Dyson ensemble is just the Unitary (U) one, with $\beta=2$.
In fact, the situation is a bit more complicated,
because of the specific symmetry of
the Bogolyubov-De Gennes equations determining
the core levels: each of the positive-energy levels $E^+_i = E >0$
has its exact mirror counterpart $E^-_i = - E$.
As a result, the
RME relevant for our problem does not coincide exactly
with Unitary or any other
of the standard Wigner-Dyson ensembles.
Fortunately, the general classification
of RME related to the mixed superconductive-normal systems
was developed very recently by Atland and Zirnbauer (AZ) \cite{AZ},
who identified 4 different types
of RME depending on the presence or
absence of time- and spin-reversal symmetries.
The class C of the AZ classification just corresponds
to our problem at hand.
The level statistics within the C class can be described via the
multi-particle ``wavefunction" of auxiliary ``free fermions"
in the same way
as it was done for the standard Unitary ensemble \cite{Alt};
the only difference is the condition of vanishing amplitude
of a single-particle ``wavefunctions"
in the origin of the energy axis $E=0$.
As a result, the level statistics within the C class coincides
with that of the U class as far as
highly excited states with $E_i \gg \hbar\omega_0$ are involved,
but differs for low-lying states.

It is very instructive to compare the dissipation rate
obtained via (\ref{C0}) with the results
of the quasiclassical kinetic equation
approach \cite{KK}.
As it will become clear soon, the applicability of the
result (\ref{C0}) to the vortex dynamics is limited
to the case when highly
excited levels are important,
so at this stage we can neglect difference between
C and U ensembles, and use (\ref{C0}) with $\beta=2$.
Comparing the definition of the vortex damping coefficient $\eta$
($\partial W/\partial t =  \eta {\bf v_v}^2$) with Eq.~(\ref{C0}),
one finds
\beq
 \pi\hbar C(0) = \eta_{Kubo} = \pi\hbar n_{2D} (\omega_0\tau),
\label{eta}
\eeq
where the second equality in (\ref{eta})
follows from the results of \cite{KK}
in the limit $\omega_0\tau \ll 1$,
$n_{2D} = n_e d$ is the areal density of electrons,
$d$ is the interlayer spacing,
and we put subscript ``Kubo" to emphasize
that the expression (\ref{eta}) has the same meaning
and the domain of applicability as the Kubo-Greenwood formula.
The damping coefficient $\eta$ is proportional
to the longitudinal flux-flow conductivity:
$\sigma_{xx} = \eta e c/\pi\hbar B$.
The parameter $C(0)$ has the dimension of inverse area;
one can define characteristic length ${\cal L} = C^{-1/2}(0)$
with the following meaning:
when vortex moves over the length ${\cal L}$,
the characteristic displacements of the core levels
inside it become of the order of the mean spacing $\omega_0$.
Then Eq.~(\ref{eta}) tells us that
${\cal L} = (n_{2D}\omega_0\tau)^{-1/2}$.
Note that ${\cal L}$ decreases
(i.e. sensitivity of the level positions
to the shift of a vortex increases)
with the decrease of disorder,
and becomes of the order of the Fermi wavelength
at $\omega_0 \tau \sim 1$, i.e. on the border  of applicability of
Eq.~(\ref{eta}).
On the other hand, in the extremely disordered limit,
$k_F l \sim 1$,
the length ${\cal L}$ would become of the order of the ``clean"
coherence length $\xi_0$.

There are two ways to understand the relation (\ref{C0}):
i) to consider open system with finite width
of energy levels and use the standard
Kubo-Greenwood approach \cite{AltSai},
and ii) to work with strictly discrete levels but
take into account the nonadiabatic transitions
between time-dependent
(due to variation of external parameter $X$)
levels $E_i(X)$, as it was done in \cite{Wilk1,Wilk2}.
The first approach is clearly invalid in our case as far as
the inelastic width $\Gamma \ll \omega_0$.
Following the second approach
\cite{Wilk2} we find that Kubo-like expression
(\ref{eta}) can be safely used if the characteristic frequency
of level perturbations $v_v/{\cal L}$ is much higher than
$\omega_0$, i.e. at $v_v \gg v_K = v_0 (k_F l)^{-1/2}$,
where $v_0 = \Delta/p_F$.
On the other hand, at low vortex
velocities $v_v \ll v_K $ and low temperatures
the probability of Zener transitions between levels
is exponentially low ($\sim \exp(- v_K/v_v)$)
when the interlevel spacing is of the order of its
average value $\hbar\omega_0$.
In this case dissipation
occurs when in course of ``level dynamics"
the spacing between some pair of levels becomes very small,
$\delta E \leq  E_Z = \hbar\omega_0 (v_v/v_K)$
and Zener tunneling becomes probable.

The crucial stage
of the dissipation process is determined by the non-adiabatic
transitions of some electron from the highest negative-energy
(filled) core level
$E_{-1}$ to the lowest positive-energy level $E_{0}$.
The rate of these
transitions can be calculated by the method
\cite{Wilk1} modified for the C class
of the AZ classification (details will
be published elsewhere \cite{long}):
\beq
R_0 = v_v \!\int\limits_0^{\infty}\! dA
          \!\int\limits_0^{\infty}\! d\epsilon N(\epsilon,A)
        \exp\! \left(\!-\frac{2\pi\epsilon^2}{\hbar A v_v} \right)
    = \frac{\eta_{Kubo}}{2\hbar\omega_0}v_v^2,\!
\label{R}
\eeq
where
$%
N(\epsilon,A) =
\frac{2\pi^{3/2}\epsilon}{ \vphantom{A^A_A} 3\hbar^3\omega_0^3}
\left[
  \frac{A^2}{ \vphantom{A^A_A} 8\hbar^2\omega_0^2 C(0)}
\right]^{3/2}
\exp
  \left(
    -\frac{A^2}{ \vphantom{A^A_A} 8\hbar^2\omega_0^2 C(0)}
  \right)%
$
is the joint probability distribution
for the energy $\epsilon$ of the lowest positive level
and its ``velocity" $A$
far from the ``avoided crossing" region.
Note that the rate $R_0$ is
just the factor 2 lower than the analogous result
for the Unitary ensemble \cite{Wilk1},
which would also be valid in our problem
if the transitions between highly exited levels
would be considered: $R_{|i| \gg 1} \approx 2R_0$.
In a similar problem treated in \cite{Wilk1}
(where all transition rates were equal, $R_i =R$),
the rate of energy absorption from the source
was given by $\hbar\omega_0 R$,
which coincided exactly with the result
for the Kubo regime, Eq.~(\ref{C0}).
This striking coincidence takes place,
as was shown in \cite{Wilk1,Wilk2},
{\it only} for the Unitary ensemble
(ensembles with the level repulsion parameter $\beta = 1, 4$ lead to
velocity-dependent friction coefficient $\eta \propto v^{\beta/2-1}$).

In our case the situation is more complicated
due to $i$-dependence of the transition rates $R_i$;
as a result, the dissipation rate
depends on the distribution function $f_i(t)$ for the
electron's population of the $i$'th core level,
which is determined by
competition between Zener processes of excitation
(the energy being absorbed from the source)
and energy relaxation to the phonon ``bath" caused
by electron-phonon interaction.
The rate of electron transitions between
the core levels with energy difference $\hbar\omega$
due to emission of 3D phonons
can be estimated \cite{long} as
$\Gamma(\omega) = \omega \gamma_{ph}$, where
$\gamma_{ph} = \nu (\omega_0\tau)\hbar\omega_0^2n_{2D}/\rho s^3 \ll 1$
(here it is assumed that $T\leq \hbar\omega$,
and $\nu$ is the numerical coefficient of order 1).
Thus the kinetic
equation for the distribution function $f_i(t)$ is:
\begin{eqnarray}
\frac{\partial f_i(t)}{\partial t}  +
    R_{i+1}(f_i-f_{i+1}) +
    R_{i} (f_i-f_{i-1}) = \nonumber \\
  - \omega_0\gamma_{ph}
\Biggl( 
    \sum\limits_{j<i} (i{-}j) f_i [1-f_j]
    - \sum\limits_{j>i} (j{-}i) f_j [1-f_i]
\Biggr), 
\label{master}
\end{eqnarray}
which is a kind of discrete ``diffusion equation"
in the energy space.
Dimensional estimates show
that the characteristic width of the stationary
solution $f^{st}_i$ is
$i_{char} \sim (R_0/\omega_0\gamma_{ph})^{1/4} \sim %
\sqrt{v_v/v_K}(\Omega_D E_F/\hbar\omega_0^3\tau)^{1/4}$;
usually $i_{char}$ is large since
$v_v$ should be larger than
the pinning-determined critical velocity $v_c$,
(estimate for $v_c/v_K$ will be given below),
whereas the second factor in the above estimate
is always very large.
One can qualitatively associate with $i_{char}$ some
``effective local temperature"
$T_{ef\!f}(v_v) \sim \hbar\omega_0 i_{char}\gg \hbar\omega_0$
of the core-localized electrons.
Due to large effective width of $f^{st}_i$,
the energy dissipation rate $\partial W/\partial t$
is close to its value $2R_0\hbar\omega_0$
for the Unitary ensemble (all $R_i = 2R_0$), which is
also the result in the Kubo regime, as mentioned above.
In order to find correction $\eta-\eta_{Kubo}$
we need to specify the rates $R_i$.
It is very difficult to determine $R_i$ for general $i \sim 1$
because of the:
i) randomness of energies where ``avoided crossing"
between levels $i-1$ and $i$ happens and
ii) energy-dependence of the mean
density of states for the C-ensemble.
Thus we employ the simplest model
with correct asymptotic behavior of rates:
$R_{i\neq 0} = R_{\infty}=2R_0$.
Within this model and at $i_{char} \gg 1$ we get
$\partial W/\partial t = 2R_0(\hbar\omega_0 + %
\langle E_0\rangle (f_0-f_{-1}))$,
which leads finally to the upper bound (cf.~below) for the vortex
damping coefficient $\eta =(\partial W/\partial t)/v_v^2$:
\beq
\frac{\eta}{\eta_{Kubo}}  \leq 1
    - \nu_1 \sqrt{\frac{v_K}{v_v}} 
\cdot\gamma_{ph}^{-1/4},
\label{eta2}
\eeq
where $\nu_1 \sim 1$.
Eq.~(\ref{eta2}) together with the relation between vortex velocity
and electric field, $v_v = cE/B$,
leads to the announced in Eq.~(\ref{new}) result
for the conductivity $\sigma_{xx}$.
Eq.~(\ref{eta2}) is valid up to temperatures $T \leq T_{ef\,f}(v_v)$;
at higher $T$ the width of $f_i^{st}$ and damping coefficient are
determined by temperature instead of vortex velocity.
Although the final result for $\sigma_{xx}$ is rather close to
the one obtained within quasiclassical picture
(where continuous spectrum of electron states is assumed), the intrinsic
mechanism  of dissipation is quite different.  One can understand it in the
following terms:  the energy dissipation rate is a product of the rate of
inelastic transitions $R_{in}$ by the characteristic amount of energy $\delta
E$ transferred at each transition; within our picture $R_{in}$ is much lower
than in the quasiclassic approach, whereas $\delta E$ is much larger by
almost the same factor.  This consideration suggests that Eq.~(\ref{new})
does not necessarily mean that the resistivity
$\rho_{xx} = \sigma_{xx}^{-1}$
as the quasiclassical results (\ref{BS}) would tell us.
The point is that Hall conductivity $\sigma_{xy}$,
being nondissipative, does not vanish
in the absence of nonadiabatic transitions and the Hall
angle $\theta_H = {\rm arctan}(\sigma_{xy}/\sigma_{xx})$
is not necessarily small even at $\omega_0\tau \ll 1$.
The calculation of $\sigma_{xy}$ within our picture
will be postponed for the future studies;
here we just note that it can't be done within purely
RMT approach, since the latter neglects completely
the correlations between
energies of the core levels and their angular momenta.
Such correlations
(which survive under moderate disorder) are
irrelevant for the energy dissipation rate
(which depends on the short-scale
structure of level correlations only),
but are crucial for the transverse
nondissipative force acting on a vortex.

We treat Eq.~(\ref{eta2})
as an upper bound for the dissipation due to the
following two reasons:
i) we have used, when deriving Eq.~(\ref{eta2}), a model that overestimates
transition rates $R_{n\neq0}$;
ii) we have employed purely classical
master equation (\ref{master}) for the
probability distribution function $f_i(t)$.
It means that we neglected
possible phase coherence of electron states on a time scale
$R_0^{-1}$ between two successive Zener transitions.
It was shown in
\cite{Gefen} that in the opposite case
of a strict phase coherence the
system exhibits a kind of 1-dimensional localization
(with the energy $E_n$ playing the role of a ``coordinate");
as a result, $(\partial W/\partial t)_{st}=0$.
We expect these phase-coherence effects
to be weak in the case of a 3D vortex
composed from ``pancakes",
since the multiplicity of allowed electron states
corresponding to each ``level" (in fact, very narrow band)
$i$ means that the problem of energy-localization
should be considered as {\it quasi}-1D,
with a rather large number of ``transverse channels".
Thus the corresponding ``localization length"
in the energy space $l_E$ is expected to be large
compared to the characteristic width $i_{char}$ of ``classical"
stationary distribution function,
so the energy-localization effects are rather weak.
However, these effects should be important for the case of a
purely 2D superconductive system.

In the above calculations we have assumed
that the vortex is moving with
a constant velocity $v_v$,
which means that the driving current density
$j = v_v e\eta/(\pi\hbar \cos\theta_H)$ is
higher than the pinning-induced critical current density $j_c$,
so the corresponding ``critical velocity" $v_c \ll v_v$.
On the other hand, $v_v$ was assumed be small compared to
$v_K = \Delta/\hbar k_F^{3/2}l^{1/2}$.
Critical current density for the pinning
of individual pancake vortex
in layered superconductor on weak impurities
can be estimated \cite{review} as
$j_c \approx j_0 (0.1\sigma_{imp}/l d)^{1/2}$,
where $j_0$ is the depairing
current density and $\sigma_{imp} = (l n_i)^{-1}$
is the electron-impurity scattering crossection
($n_i$ is the impurity concentration).
Using ``pessimistic" estimate $\theta_H \ll 1$
we get the compatibility condition
for our theory in the form
$5\cdot 10^{-2} k_F\sigma_{imp}/d \ll (\omega_0\tau)^2 \leq 1$.
Thus the proposed scenario
may be realized in the case of very weak impurities
with a crossection small on atomic scale.
Note however that the above
condition may actually be less stringent
due to the decrease of $j_c$ with the
magnetic field  already in the range $B \ll H_{c2}$ \cite{review},
and also in the presence of a considerable
Hall component of motion, $\tan\theta_H \geq 1$.
We believe that
the necessary conditions may be fulfilled
in the clean single crystals of layered
superconductor NbSe$_2$ where the main source of disorder
is due to random substitution of small
percentage of Nb by Ta,
which is known to produce very weak impurity centers
\cite{shobo}.
Our results may also be relevant for layered HTSC,
however in that case the role of
$d$-wave pairing symmetry should be studied.

In conclusions,
we proposed new mechanism of flux-flow dissipation
operative in layered superconductors at low temperatures.
Dissipative conductivity $\sigma_{xx}$ is found to be close to the
classical results \cite{BS,KK} but is slowly field-dependent.
We suggest that, under the same conditions,
the Hall conductivity $\sigma_{xy}$ and the
flux-flow resistivity $\rho_{xx}$ may differ considerably
from their classical values.
Useful discussions with B.L.Altshuler, S.Bhattacharya,
V.B.Geshkenbein, N.B.Kopnin, V.E.Kravtsov, A.I.Larkin,
V.M.Marikhin, V.P.Mineev and G.E.Volovik
are gratefully acknowledged.
This research was supported by the RFBR grant \# 95-02-05720,
DGA grant \# 94-1189 (M.V.F.) and ISSEP\- grant \# a96-568 (M.A.S.).

\end{multicols}

\end{document}